\documentclass[twoside,twocolumn,9pt]{article}
\usepackage{extsizes}
\usepackage[super,sort&compress,comma]{natbib} 
\usepackage[version=3]{mhchem}
\usepackage[left=1.5cm, right=1.5cm, top=1.785cm, bottom=2.0cm]{geometry}
\usepackage{balance}
\usepackage{mathptmx}
\usepackage{sectsty}
\usepackage{graphicx} 
\usepackage{lastpage}
\usepackage[format=plain,justification=justified,singlelinecheck=false,font={stretch=1.125,small,sf},labelfont=bf,labelsep=space]{caption}
\usepackage{float}
\usepackage{fancyhdr}
\usepackage{fnpos}
\usepackage[english]{babel}
\addto{\captionsenglish}{%
  
}
\usepackage{array}
\usepackage{droidsans}
\usepackage{charter}
\usepackage[T1]{fontenc}
\usepackage[usenames,dvipsnames]{xcolor}
\usepackage{setspace}
\usepackage[compact]{titlesec}
\usepackage{hyperref}
\usepackage[utf8]{inputenc}
\usepackage{gensymb}
\usepackage{caption}
\usepackage{subcaption}
\usepackage{epstopdf}%This line makes .eps figures into .pdf - please comment out if not required.

\definecolor{cream}{RGB}{222,217,201}

\begin{document}

\pagestyle{fancy}
\thispagestyle{plain}
\fancypagestyle{plain}{
%%%HEADER%%%
\renewcommand{\headrulewidth}{0pt}
}
%%%END OF HEADER%%%

%%%PAGE SETUP - Please do not change any commands within this section%%%
\makeFNbottom
\makeatletter
\renewcommand\LARGE{\@setfontsize\LARGE{15pt}{17}}
\renewcommand\Large{\@setfontsize\Large{12pt}{14}}
\renewcommand\large{\@setfontsize\large{10pt}{12}}
\renewcommand\footnotesize{\@setfontsize\footnotesize{7pt}{10}}
\makeatother

\renewcommand{\thefootnote}{\fnsymbol{footnote}}
\renewcommand\footnoterule{\vspace*{1pt}% 
\color{cream}\hrule width 3.5in height 0.4pt \color{black}\vspace*{5pt}} 
\setcounter{secnumdepth}{5}

\makeatletter 
\renewcommand\@biblabel[1]{#1}            
\renewcommand\@makefntext[1]% 
{\noindent\makebox[0pt][r]{\@thefnmark\,}#1}
\makeatother 
\renewcommand{\figurename}{\small{Fig.}~}
\sectionfont{\sffamily\Large}
\subsectionfont{\normalsize}
\subsubsectionfont{\bf}
\setstretch{1.125} %In particular, please do not alter this line.
\setlength{\skip\footins}{0.8cm}
\setlength{\footnotesep}{0.25cm}
\setlength{\jot}{10pt}
\titlespacing*{\section}{0pt}{4pt}{4pt}
\titlespacing*{\subsection}{0pt}{15pt}{1pt}
%%%END OF PAGE SETUP%%%

%%%FOOTER%%%
\fancyfoot{}
\fancyfoot[LO,RE]{\vspace{-7.1pt}\includegraphics[height=9pt]{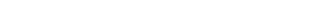}}
\fancyfoot[CO]{\vspace{-7.1pt}\hspace{13.2cm}\includegraphics{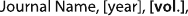}}
\fancyfoot[CE]{\vspace{-7.2pt}\hspace{-14.2cm}\includegraphics{head_foot/RF}}
\fancyfoot[RO]{\footnotesize{\sffamily{1--\pageref{LastPage} ~\textbar  \hspace{2pt}\thepage}}}
\fancyfoot[LE]{\footnotesize{\sffamily{\thepage~\textbar\hspace{3.45cm} 1--\pageref{LastPage}}}}
\fancyhead{}
\renewcommand{\headrulewidth}{0pt} 
\renewcommand{\footrulewidth}{0pt}
\setlength{\arrayrulewidth}{1pt}
\setlength{\columnsep}{6.5mm}
\setlength\bibsep{1pt}
%%%END OF FOOTER%%%

%%%FIGURE SETUP - please do not change any commands within this section%%%
\makeatletter 
\newlength{\figrulesep} 
\setlength{\figrulesep}{0.5\textfloatsep} 

\newcommand{\topfigrule}{\vspace*{-1pt}% 
\noindent{\color{cream}\rule[-\figrulesep]{\columnwidth}{1.5pt}} }

\newcommand{\botfigrule}{\vspace*{-2pt}% 
\noindent{\color{cream}\rule[\figrulesep]{\columnwidth}{1.5pt}} }

\newcommand{\dblfigrule}{\vspace*{-1pt}% 
\noindent{\color{cream}\rule[-\figrulesep]{\textwidth}{1.5pt}} }

\makeatother
%%%END OF FIGURE SETUP%%%

%%%TITLE, AUTHORS AND ABSTRACT%%%
\twocolumn[
  \begin{@twocolumnfalse}
{\includegraphics[height=30pt]{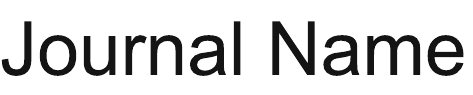}\hfill\raisebox{0pt}[0pt][0pt]{\includegraphics[height=55pt]{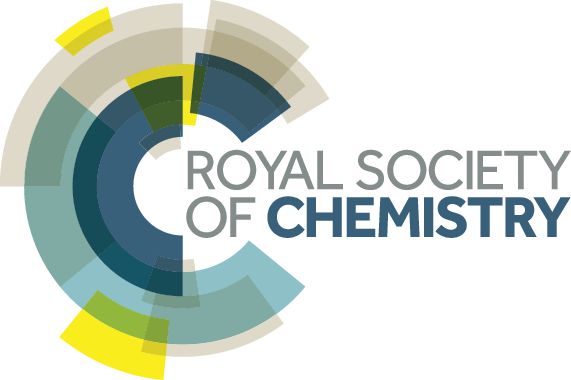}}\\[1ex]
\includegraphics[width=18.5cm]{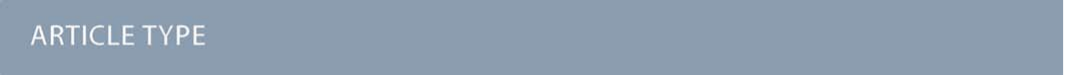}}\par
\vspace{1em}
\sffamily
\begin{tabular}{m{4.5cm} p{13.5cm} }

\includegraphics{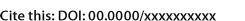} & \noindent\LARGE{\textbf{From Femtoseconds to Gigaseconds: The SolDeg Platform for the Performance Degradation Analysis of Silicon Heterojunction Solar Cells}} \\
\vspace{0.3cm} & \vspace{0.3cm} \\

 & \noindent\large{Davis Unruh,$^\ast$\textit{$^a$} Reza Vatan Meidanshahi,$^b$ Chase Hansen,$^a$ Salman Manzoor,$^b$ Stephen M. Goodnick,$^b$ Mariana I. Bertoni,$^b$ and Gergely T. Zimanyi$^a$} \\

\includegraphics{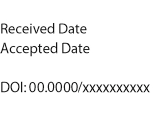} & \noindent\normalsize{Heterojunction Si solar cells exhibit notable performance degradation. We modeled this degradation by electronic defects getting generated by thermal activation across energy barriers over time. To analyze the physics of this degradation, we developed the SolDeg platform to simulate the dynamics of electronic defect generation. First, femtosecond molecular dynamics simulations were performed to create a-Si/c-Si stacks, using the machine-learning-based Gaussian approximation potential. Second, we created shocked clusters by a cluster blaster. Third, the shocked clusters were analyzed to identify which of them supported electronic defects. Fourth, the energy barriers that control the generation of these electronic defects and their distribution were determined. Fifth, an accelerated Monte Carlo method was developed to simulate the time dependent defect generation across the barriers by thermal activation. -- Our main conclusions are as follows. (1) The degradation of a-Si/c-Si heterojunction solar cells via defect generation is controlled by a very broad distribution of energy barriers. (2) We developed the SolDeg platform to track the microscopic dynamics of defect generation across this wide barrier distribution, and determined the time dependent defect density $N(t)$ from femtoseconds to gigaseconds, over 24 orders of magnitude in time. (3) We have shown that a simple, stretched exponential analytical form can successfully describe the defect generation $N(t)$ over at least ten orders of magnitude in time. (4) We found that in relative terms $V_\mathrm{oc}$ degrades at a rate of 0.2\%/year over the first year. This rate slows with advancing time. (5) We developed the Time Correspondence Curve to calibrate and validate the accelerated testing of solar cells. We found a compellingly simple scaling relationship between accelerated and normal times $t_\mathrm{accelerated} \propto t_\mathrm{normal}^{0.85}$, which can be used to calibrate accelerated testing protocols. (6) We ourselves also carried out experimental studies of defect generation in a-Si:H/c-Si HJ stacks. We found a relatively high degradation rate at early times, that slowed considerably at longer time scales.} \\

\end{tabular}

 \end{@twocolumnfalse} \vspace{0.6cm}

]
%%%END OF TITLE, AUTHORS AND ABSTRACT%%%

%%%FONT SETUP - please do not change any commands within this section
\renewcommand*\rmdefault{bch}\normalfont\upshape
\rmfamily
\section*{}
\vspace{-1cm}

%%%FOOTNOTES%%%
\footnotetext{$^a$Physics Department, University of California, Davis, Davis, CA, 95616, USA. Email: dgunruh@ucdavis.edu}
\footnotetext{$^b$School of Electrical, Computer and Energy Engineering, Arizona State University, Tempe, AZ, 85287-5706, USA}
%%%END OF FOOTNOTES%%%

%%%MAIN TEXT%%%%

% \begin{description}

% \item[PACS numbers] 73.63.Kv
% %https://publishing.aip.org/publishing/pacs/pacs-2010-regular-edition

% \item[Keywords] Heterojunctions, Defects, Degradation, Molecular dynamics, Density functional theory
% \end{description}

\section{Introduction}
\subsection{Solar Cell Degradation}

Heterojunction (HJ) Si solar cells have world record efficiencies approaching 27\%, due to the excellent surface passivation by their amorphous Silicon (a-Si) layer that leads to low surface recombination velocities and high open circuit voltages $V_{OC}$. In spite of the impressive efficiency records, HJ Si cells have not yet been widely adopted by the market because of the perceived challenge that HJ cells may exhibit accelerated performance degradation, possibly related to their a-Si layer. Traditional crystalline Si (c-Si) modules typically exhibit about a 0.5\%/yr efficiency degradation, primarily via their short circuit current $I_{sc}$ and the fill factor FF, typically attributed to external factors, such as moisture ingress and increased contact resistance. In contrast, in 2018 two papers reported studies of the degradation of fielded Si HJ modules over 5-10 years \cite{DelineChris2018SHSF,ChoiSungwoo2018ADRo}. They reported degradation rates close to 1\%/yr, about twice the rate of traditional cells. These papers pointed to a new degradation channel, the decay of $V_{OC}$, at a rate of about 0.5\%/yr. The decay of $V_{OC}$ suggests that the degradation is possibly due to internal factors, increasing recombination either at the a-Si/c-Si interface, or in the a-Si layer. Such increased recombination is typically caused by the increase of the electronic defect density.

These initial reports on fielded panels were followed up by in-laboratory analysis. The Bertoni group has studied the surface recombination velocity (SRV) at the a-Si/c-Si interface in HJ stacks. By applying a model for the recombination at the a-Si/c-Si interface to their temperature- and injection-dependent SRV data, they analyzed the degradation of the carrier lifetime and were able to attribute it to a loss of chemical passivation \cite{BertoniMarianaI.2018IitD}. More recently, Holovsky et al. investigated ultrathin layers of hydrogenated amorphous silicon (a-Si:H), passivating the surface of crystalline silicon (c-Si) \cite{DeNicolasSilviaMartin2020ASIS}. These authors applied highly sensitive attenuated total reflectance Fourier-transform infrared spectroscopy, combined with carrier lifetime measurements. They manipulated the a-Si/c-Si interface by applying different surface, annealing, and aging treatments. Electronic interface properties were discussed from the perspective of hydrogen mono-layer passivation of the c-Si surface and from the perspective of a-Si:H bulk properties. They concluded that both models have severe limitations and called for a better physical model of the interface \cite{DeNicolasSilviaMartin2020ASIS}.

Understanding the degradation of the passivated c-Si surface is important not only for understanding a-Si/c-Si heterojunction solar cells. The PV industry roadmap shows that among newly installed modules, the fraction of advanced Passivated Emitter/Rear Contact (PERC) modules will rapidly rise above 50\% in the next 3 years. One of the advanced features of these PERC cells is the improved interface passivation with the application of elevated levels of hydrogen. However, the increased efficiency was accompanied by notable levels of degradation
\cite{HamerPhillip2020HdEt,TongHongbo20202emi,HamerPhillipG.2018HidA}. By experiments and by including all three charge states of hydrogen in their modeling, the authors speculated that the PERC cell degradation both in the dark and under illumination could be explained by the migration of and interaction between hydrogen ions in different charge states. 

To summarize, the accelerated degradation of $V_{OC}$ slows the market acceptance of the world-efficiency-record holder HJ Si modules, and impacts the introduction of the advanced PERC cells, thereby impacting the entire PV industry roadmap. Therefore, analyzing and mitigating this degradation process is of crucial importance. 

\subsection{Defects in Amorphous Si}
Photoinduced degradation of a-Si under prolonged exposure to intense light was first studied, measured and modeled by Staebler and Wronski \cite{staeblerwronski}. They reported that the degradation is characterized by a remarkably universal $t^{1/3}$ power-law temporal growth of the defect density. This behavior has become known as the Staebler-Wronski effect (SWE). 

The  SWE was analyzed by different methods. Some groups performed electron spin resonance (ESR) measurement on a-Si (a-Si:H) to experimentally detect the increase of the density of dangling bonds induced by light exposure \cite{dersch,stutzmann,SchneggA.2015Saep}. Some of these papers also developed a phenomenological model to predict the SW defect-increase as a function of exposure time and light intensity. Other groups used the photocurrent method (PCM) to detect the change of defect density of a-Si under light exposure. In agreement with ESR experiments, PCM also revealed the increase of defect density under light exposure. While the ESR and PCM defect density measurements yielded analogous results, it is recalled here that they capture different type of defect states \cite{Shimizu_2001, Shimizu_2002}. ESR detects all neutral defect states that only include dangling bonds (DBs), while PCM detects both neutral and charged defect states that include DBs and other types of defect states. Therefore PCM measurements revealed that the origin of defect states might be a result of different type of general structural disorders beyond DBs.\cite{gotoh, Shimizu_1997} 

Recently, Wronski argued that three distinct defect states, A/B/C, are needed to account for all the data, instead of the standard single “midgap dangling bond” defect \cite{NiuXinwei2014TLRo}. The A/B states are efficient electron recombination centers, while the C states recombine holes efficiently. Wronski speculated that these states are differentiated by their different structures: dangling bonds, mono- and divacancies, as also advocated by Smets. Other groups also analyzed their data in terms of three distinct states \cite{BertoniMarianaI.2018IitD,TongHongbo20202emi}. However, they focused on the alternative picture that the defect states may be the three charge states H+, H0, and H- of hydrogen. In addition to these experimental works, recent theoretical and computational papers also analyzed the defect states in a-Si, and they concluded that besides dangling bonds, highly strained bonds also contribute to midgap states significantly \cite{khomyakov,Reza}.

To summarize, while a fair amount of progress has been achieved in characterizing defect generation in a-Si, its underlying mechanism and connection to the different types of structural disorder and defects is far from being settled and understood. The problem is still open to question. For this reason, in this paper we analyze the above problem of defect generation in a-Si/c-Si heterojunction solar cells, with a possible relevance for PERC cell passivation.

\section{Methods and Results}
\subsection{The SolDeg Platform}

To address the above-described $Sol$ar cell $Deg$radation, we have developed the SolDeg platform to model electronic defect generation in a-Si/c-Si heterojunctions, which consists of the following hierarchical stages. (1) Creating a-Si/c-Si stacks. (2) Generating shocked clusters as likely hosts of electronic defects. (3) Identifying shocked clusters that actually host electronic defects. (4) Determining the energy barriers that control the generation of these electronic defects; and determining their distribution. (5) Determining the temporal evolution of the defect density from the energy barrier distribution. This SolDeg platform is described next in detail.

\subsection{Creating the Amorphous/Crystalline Si Stacks: Machine-Learning Driven Molecular Dynamics}

The SolDeg platform starts with creating a-Si/c-Si heterojunction structures, or stacks. We first created pure a-Si structures, which were carefully optimized in order to match lab-grown a-Si as closely as possible. Second, we placed these optimized a-Si structures on top of slabs of c-Si, and then annealed the interface region. This approach was chosen in order to create the most realistic a-Si atomic structures possible, while still yielding a reasonable aSi/cSi interface region. The details of this approach are as follows. 

To create pure amorphous Si structures, one performs melt-quench molecular dynamics (MD) simulations. In a melt-quench MD simulation, a crystalline Si structure is heated past its melting point to generate liquid Si, which is then quenched down to low temperatures at an appropriate rate. This method is widely used for generating amorphous Si networks. It is known that the choice of the interatomic potential used for these MD simulations has a substantial effect on the results. Classical parametric interatomic potentials, such as the Tersoff \cite{Tersoff} and Stillinger-Weber (SW) \cite{SW} potentials, have a limited number of parameters/descriptors, and are typically fitted against experimental structural data under a specific set of conditions such as a particular material composition and temperature range. As such, their accuracy in reproducing a wider variety of structural properties of the specific material, or in simulating different temperature ranges or material structures than they were fitted to, often limits the precision of the results. 

For example, the excess energy (energy compared to diamond-type Si) of the a-Si resulting from melt-quench simulations performed with these interatomic potentials is typically $>0.20$ eV/atom, falling outside of the lab-grown a-Si excess energy range of $0.07-0.15$ eV/atom \cite{excessenergy1,excessenergy2,excessenergy3}. The defect densities, e.g. the density of dangling and floating bonds, also differ from typical lab-grown a-Si data. MD simulations with these interatomic potentials are also unable to reach DFT-level accuracy in determining elastic constants and defect formation energies. For all the above reasons, the MD-created a-Si systems need to be further optimized by Density Functional Theory (DFT)\cite{Reza}. However, the need to use DFT slows down the computational time substantially, and thus limits the accessible system sizes substantially.

To improve the accuracy of our MD simulations in all aspects compared to using these standard interatomic potentials, we instead adopted a Machine-Learning driven general-purpose interatomic potential which has been created for Si \cite{PRX}. This machine-learning driven approach uses the framework of the Gaussian approximation potential (GAP) with a smooth overlap of atomic positions (SOAP) kernel, and has been specifically developed so that GAP-based MD simulations yield DFT-level accuracy, even though they are 10x more efficient, thus enabling the faster simulation of larger systems \cite{GAP1, GAP2}. Hereafter, we will refer to this potential simply as the Si GAP. It has been shown that the Si GAP captures more than a dozen experimentally measured quantities significantly better than any of the other available interatomic potentials \cite{PRX}. We show below that adopting the Si GAP for our MD simulations yield superior a-Si structures after only a minimum level of DFT optimization. Now we proceed with the technical simulation details.

We generated the a-Si/c-Si stacks by MD simulations, carried out using the LAMMPS software package \cite{LAMMPS}. The simulation time step was 1 fs. Our melt-quench simulations started with crystalline Si cubic supercells containing 216 Si atoms, with three dimensional (3D) periodic boundary conditions. The lattice constant $a_0$ was chosen to be $5.43\,$\AA, and the dimensions of the supercell $a=b=c=3a_0$. This lattice constant was chosen to ensure that the mass density of the resulting a-Si structures was 1-3\% lower than the mass density of corresponding c-Si structures, consistent with the mass density measured by experiments on a-Si/c-Si stacks. 

The crystalline Si was first heated to 1800K to yield liquid Si. The liquid Si was subsequently re-solidified by cooling down to 1500K at a rate of $10^{13} \,\mathrm{K/s}$ before being equilibrated at 1500K for 100 ps. This solid Si was quenched further down to 500K at a rate of $10^{12} \,\mathrm{K/s}$, following previous studies \cite{PRX, cooling1, cooling2, cooling3}. The first quench was performed in the constant-volume and variable-pressure (NVT) ensemble, while the second quench was performed in the variable-volume and constant-pressure (NPT) ensemble with fixed $x$ and $y$ cell-dimensions (to match the dimensions of the c-Si unit cell in the later steps), both using a Nos\'e-Hoover thermostat and barostat. We minimized the structural energy using a GAP-driven Hessian-free truncated Newton (HFTN) algorithm to relax all atomic positions into their local minima.

These relaxed a-Si structures were further optimized with DFT, specifically making use of the Quantum Espresso 6.2.1 software package \cite{QE1, QE2}. We used the Broyden-Fletcher-Goldfarb-Shanno (BFGS) quasi-newton algorithm, based on the trust radius procedure, as the optimization algorithm.

The Perdew-Burke-Ernzerhof (PBE) exchange-correlation functional \cite{PBE} was used in both the ionic relaxation and the electronic structure calculations using periodic boundary conditions. The core and valence electron interactions were described by the Norm-Conserving Pseudopotential function. Unless otherwise stated, an energy cutoff of $12\,$Ry was employed for the plane-wave basis set and a 2×2×2 k-point mesh was used with the Monkhorst-Pack grid method for the Brillouin-zone sampling in all the calculations. Methfessel-Paxton smearing \cite{methfesselpaxton} of width $0.05\,$Ry was applied to determine the band occupations and electronic density of states.

\begin{figure}[b!]
\begin{center}
\includegraphics[width=\columnwidth]{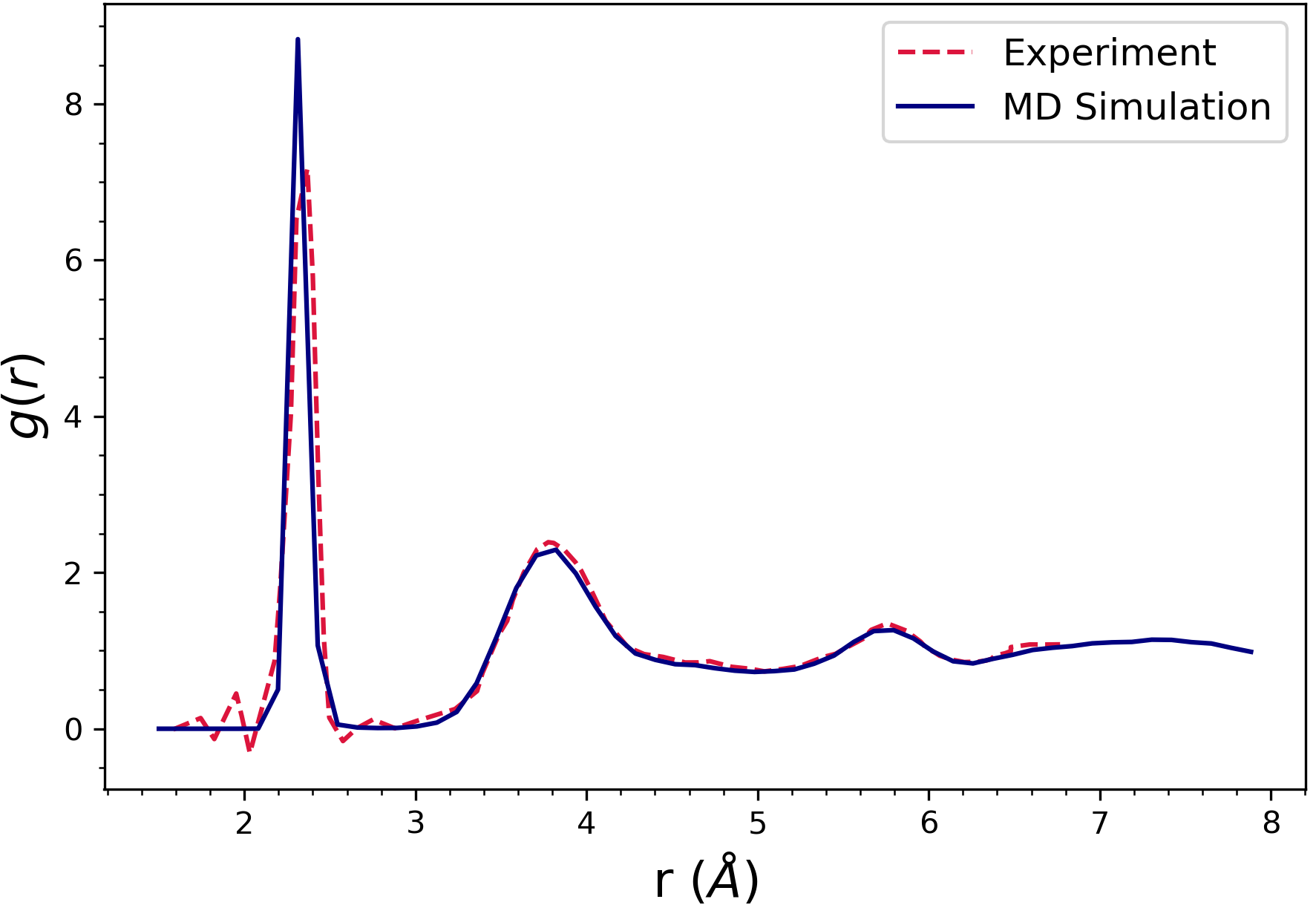}
\end{center}
\caption{Radial distribution function $g(r)$ characterizing a typical melt-quench MD a-Si structure, plotted against the experimental values of ref. \citenum{Laaziri}}
\label{RDF}
\end{figure}

Motivated by Pedersen et al., we use the excess energy, the bond angle distribution, and the radial distribution function (RDF), as the most compelling criteria to validate our generated a-Si structures against a-Si experiments \cite{Pedersen_2017}. In our structures the typical excess energies were around $0.13-0.14\,$eV/atom, well within the experimentally acceptable range of $0.07-0.15$ eV/atom. These remarkably low excess energies strongly validate the superiority of the Si GAP over traditional potentials, which yield excess energies above $0.20$ eV/atom. The bond-angle distribution was centered at 109.1$\degree$ with a width of $\pm 10.5\degree$. These values are also consistent with typical experimental values \cite{fortner}. The average Si-Si bond length was $2.38\,$\AA $\,\pm\, .04\,$\AA. Assuming a Si-Si bond-length cutoff of $2.58\,$\AA, slightly less than 10\% longer than the average bond-length, the average number of dangling bonds in each supercell was 2.2, and the average number of floating bonds was 0.8. Dangling (floating) bonds are missing (extra) bonds of a Si atom relative to the standard number of 4. Finally, the structures were further validated by calculating the radial distribution function (RDF). The RDF measures the probability of finding the center of an atom at a given distance from the center of another atom as a function of their radial separation. For a-Si, the typical RDF exhibits a strong peak centered at $2.3\,$\AA, and two weak peaks centered around $3.8\,\textrm{\AA}$ and $5.4\,\textrm{\AA}$ (the next-nearest and next-next-nearest neighbor distances in c-Si). As shown in Fig. 1, the RDFs calculated from our GAP-MD generated structures track the experimental data compellingly.

\begin{figure}[t]
\begin{center}
\includegraphics[width=\columnwidth]{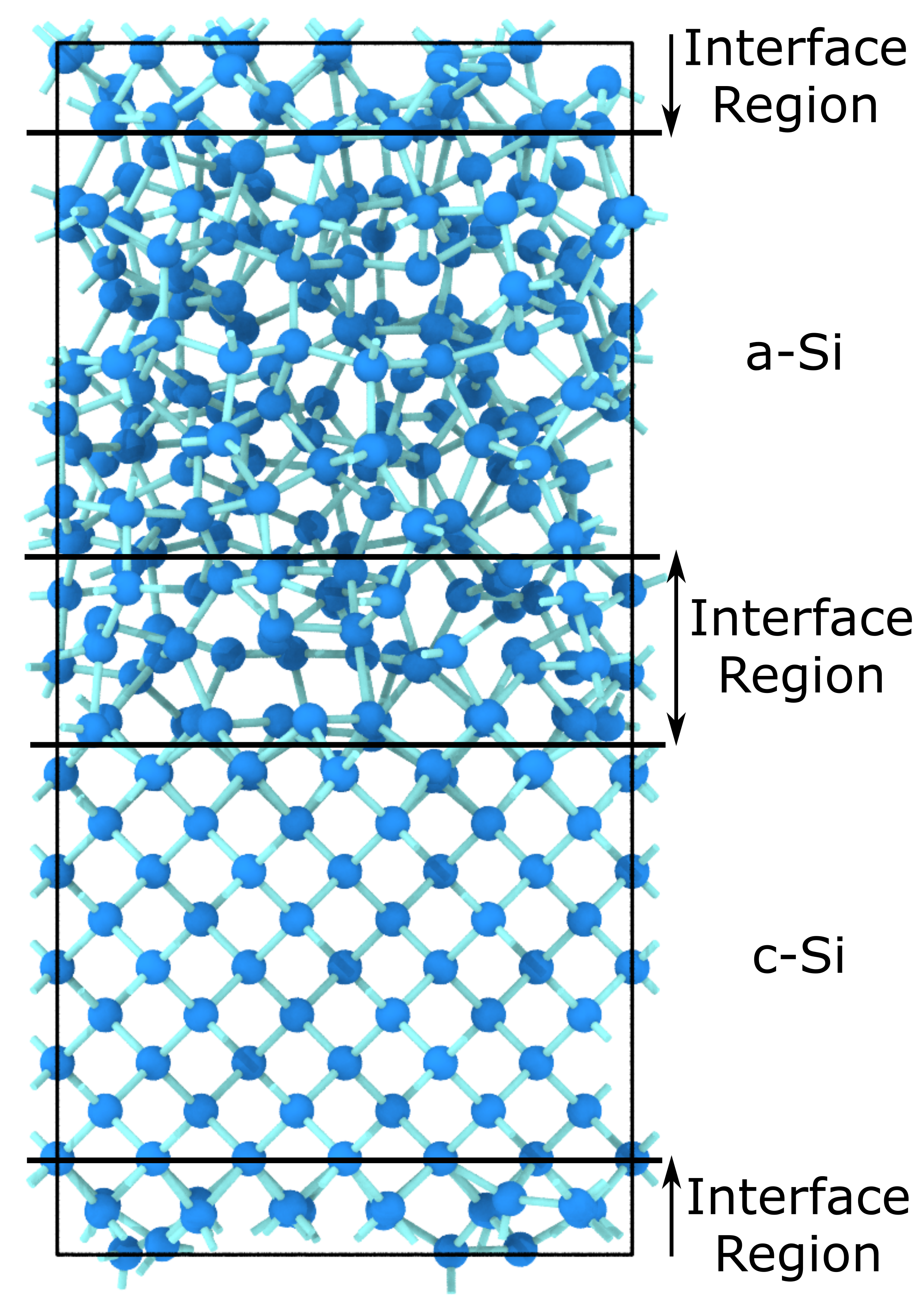}
\end{center}
\caption{Rendering of a simulated Si-heterojunction structure. Periodic boundary conditions result in the presence of two distinct interface regions.}
\label{heterojunction}
\end{figure}

We created the Si-heterojunction structures by placing the DFT-optimized a-Si on top of c-Si slabs (of the same dimensions and number of atoms as the a-Si structures). See Fig.~\ref{heterojunction}. Note that requiring periodic boundary conditions for the a-Si/c-Si stacks forces two a-Si/c-Si interfaces into the structure, as shown. At both interfaces, the a-Si was placed $1.36\,\textrm{\AA}$ ($a_0/4$) away from the edge of the c-Si slab. This distance was chosen by calculating the total energy of a series of structures where this distance was systematically varied, and choosing the distance which yielded the lowest total energy.

The resulting a-Si/c-Si interfaces are highly strained. For this reason, we relax each interface via thermal annealing. To avoid altering the structure of the carefully optimized a-Si layers, we only annealed a strip of width $a_0$ centered symmetrically at each a-Si/c-Si interface. The annealing was performed at 450K for 25 ps, and was followed by cooling down to 270K at a rate of $10^{13} \,\mathrm{K/s}$. Both steps were performed in the NVT ensemble with a timestep of 1 fs. In total, we created 50 a-Si/c-Si structures.

\subsection{Defect Generation: Shocked Cluster Generation by the Cluster Blaster}
Our overarching theory is that the HJ cell performance degradation is driven by the generation of electronic defects that act as recombination centers. We further posit that most of the electronic defects are generated by a small group of Si atoms in the a-Si transitioning from their moderately disordered cluster into a highly disordered cluster by thermal activation over an energy barrier. The transition into this highly disordered cluster can strain or break the Si bonds, thereby creating electronic defects, such as strained or dangling bonds. 

We decided to create highly disordered, "shocked" clusters by heating the cluster very quickly to excessive temperatures, followed by a comparably quick cooling: a procedure we refer to as the "cluster blaster". Using LAMMPS\cite{LAMMPS} and the ML-based Silicon Gaussian Approximation Potential (GAP) \cite{PRX}, described in the previous section, we blasted clusters of 5 atoms in our a-Si/c-Si stack, centered at the crystalline/amorphous interface to a temperature of $T=5000$K while keeping the rest of the structure frozen. We chose $T=5000$K, because we found that temperatures significantly below this value were not efficient at generating electronic defects in our systems, to be described below. We allowed the shocked clusters to evolve at this elevated temperature for 20 ns so that they could explore their configuration space extensively. After 20 ns, the shocked clusters were quenched quickly, so that they could not escape whichever highly disordered metastable configuration they were nearest to. We then performed a Hessian-free truncated Newton optimization of the quenched shocked clusters, again using the Si GAP. This cluster blaster process was repeated at the interfaces of all of our 50 a-Si/c-Si stacks at about 30 different locations each, eventually creating about 1,500 shocked clusters with the cluster blaster. 

\subsection{Defect Generation: Analysis of the Shocked Clusters for Electronic Defects}

The cluster blaster does not always induce electronic defects in the shocked clusters. To identify which cluster blasting induced electronic defects as well, the next stage of SolDeg is to measure the orbital localization of the electronic states in the a-Si/c-Si stacks with shocked clusters.

We determined the localization of the Kohn-Sham electronic orbitals in the a-Si/c-Si structures before and after the cluster blasting by using the inverse participation ratio (IPR) method. The IPR for an eigenstate $\Psi_n$ is given as:
\begin{equation}
IPR_n=\frac{\Sigma_{i=1}^I{a_{ni}^4}}{({\Sigma_{i=1}^I{a_{ni}^2}})^2}
\label{IPR-Eq}
\end{equation}
\indent where $a_{ni}$ is the coefficient of $i^{th}$ basis set orbital in $n^{th}$ Kohn-Sham orbital $\Psi_n$ ($\Psi_n=\Sigma_{i=1}^I{a_{ni}\phi_i}$) and \textit{I} is the total number of basis set orbitals used in the DFT calculation. The higher the IPR, the higher the degree of localization. The IPR for a state extended equally over all atoms is close to zero (~ O(1/N)), and for a state completely localized state on only one atom is one.

\begin{figure}[t]
\begin{center}
\includegraphics[width=\columnwidth]{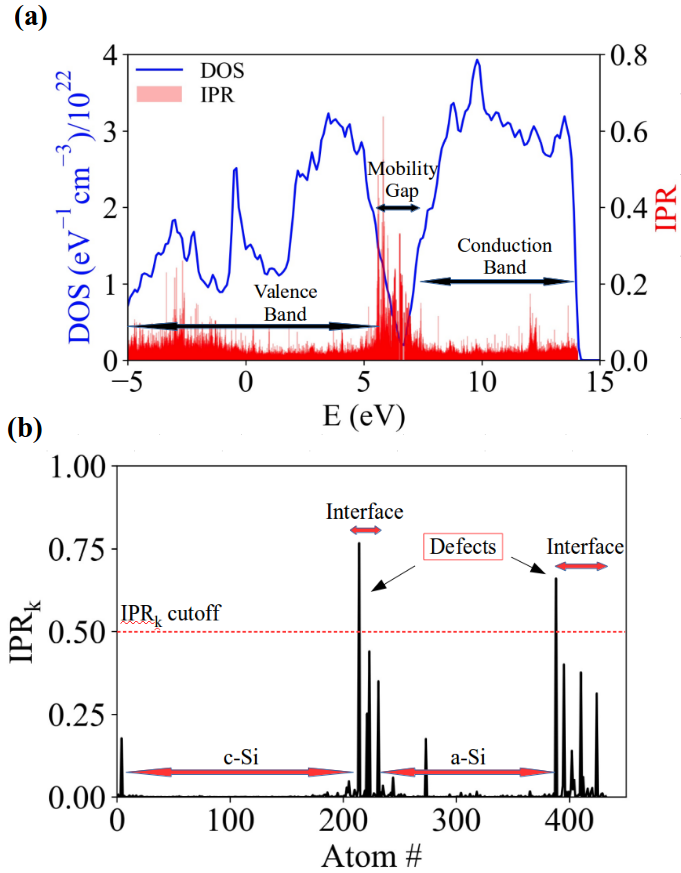}
\end{center}
\caption{(a) $IPR$ and (b) $IPR_k$ of typical a-Si/c-Si structures. }
\label{ipr}
\end{figure}

Fig.~\ref{ipr}(a) shows the IPRs, calculated for all Kohn-Sham orbitals obtained by DFT as a function of their energy for a typical a-Si/c-Si stack. Visibly, the majority of the electronic states are localized in the energy region of 5.5-7 eV, lying between the conduction band and the valence band. This region can be identified as a mobility gap because the electronic states are localized within. In contrast, the majority of the states are delocalized inside the conduction and valence bands. It is recalled that the mobility gap often differs somewhat from the density of states (DOS) gap, since the electronic states in the tails of the conduction and valence bands can be localized, separated from the delocalized continuum of band states by a "mobility edge". 

In order to determine the localization of the electronic orbitals more  we rearrange Eq. \ref{IPR-Eq} as follows:

\begin{equation}
IPR(\Psi_n)=\frac{\Sigma_{k=1}^K\Sigma_{j=1}^J{a_{nkj}^4}}{(\Sigma_{k=1}^K{\Sigma_{j=1}^J{a_{nkj}^2}})^2}
\label{IPRk-Eq}
\end{equation}

\indent where $a_{nkj}$ is the coefficient of $j^{th}$ atomic orbital belonging to the $k^{th}$ atom in the $n^{th}$ Kohn-Sham orbital. \textit{J} is the total number of atomic orbitals used in DFT calculations, which belong only to the $k^{th}$ atom in the supercell, and \textit{K} is the total number of atoms inside the supercell. We introduced the concept of Eq. \ref{IPRk-Eq} because it is capable of identifying not only that an electronic state is localized, but the location of the atom where it is localized as well by defining a quantity $IPR_{nkj}$ as follows: \cite{Reza}

\begin{equation}
IPR_{nkj}=\frac{a_{nkj}^4}{(\Sigma_{k=1}^K{\Sigma_{j=1}^J{a_{nkj}^2}})^2}
\label{IPRnki-Eq}
\end{equation}

Here $IPR_{nkj}$ is the contribution of the $k^{th}$ atom through its $j^{th}$ atomic orbital in the localization of the $n^{th}$ Khon-Sham orbital. One notes that the denominator of Eq. \ref{IPRnki-Eq} is the same for the all $IPR_{nkj}$ for a given \textit{n}. Thus, the number of $IPR_{nkj}$ values for a given \textit{k} atom for each Kohn-Sham orbital is \textit{J}. Denoting the number of Kohn-Sham orbitals as \textit{N}, each atom in the supercell has NJ $IPR_{nkj}$ values. In order to assign only one IPR value to each atom \textit{k}, we choose the maximal $IPR_{nkj}$ from among the NJ $IPR_{nkj}$ values for a fixed \textit{k}. We name this maximal value $IPR_{k}$:

\begin{equation}
IPR_k=MAX_{n,j}^k\{IPR_{nkj}\}
\label{IPRK-Eq}
\end{equation}

Fig.~\ref{ipr}(b) shows $IPR_{k}$ for a typical a-Si/c-Si structure, as a function of the atom number \textit{k}, approximately translating into the z-coordinate of the atoms. As expected, almost all of the localized states are located at the interfaces. There are no localized states in the c-Si, and only one localized state in the a-Si. This localized state distribution is reasonable given the high degree of strain at the interface, in contrast to the low strain in the a-Si, and minimal strain in the c-Si. We identify electronic states as genuine electronic defects as long as they are mostly localized on a single atom, This is captured by their $IPR_k$ value exceeding a threshold which we take as 0.5. With this threshold convention, visibly there is only one defect at each interface in Fig.~\ref{ipr}(b).

Once the IPR calculations have been completed, we can determine whether electronic defects have been successfully created in the shocked clusters by the cluster blaster. As somewhat of a surprise, we found that quite often the cluster blasting in fact did the opposite: it annealed out an already existing electronic defect instead of creating one. Therefore, we broadened the scope of our search to identify pairs of initial and final states of the cluster blasting in which the number of electronic defects differed by precisely one. This protocol picked up both the creation and the annihilation of electronic defects in the shocked clusters. We chose to only track initial-final state pairs that differed by a single defect to avoid the need of tracking defect-defect interactions that may affect our results.

\subsection{Determining Energy Barriers and Their Distribution with the Nudged Elastic Band Method}

In the next stage of SolDeg, we determined the energy barriers that control the creation and annihilation of the identified electronic defects because thermal activation across these barriers controls the temporal increase of the overall electronic defect density in a-Si/c-Si stacks, aged in the dark. We will return to light-induced defect generation in a later paper.

Once pairs of initial and final states have been identified where a single electronic defect was either created or annihilated, we employed the nudged elastic band (NEB) method \cite{NEB1, NEB2, NEB3, NEB4} to determine the energy barrier heights between these initial and final states. The nudged elastic band method connects two different local energy minima with several intermediate replica states, each connected to its nearest state neighbor with a "spring" that is nudged perpendicular to the path through state space to allow the "band" to find a saddle point. The NEB method is a standard tool for determining minimum energy paths between states in some fields, but to our knowledge, the NEB method has not been used in the solar field yet, so we will describe the method in some detail here.

NEB starts with an initial guess of a sequence of intermediate "replica" states between an initial and a final state. NEB then postulates an abstract "spring" between the adjacent replica states, to generate a tendency for sequence of replica states to evolve towards a compact, possibly lower energy path between the initial and final state.

The NEB method uses two force components to cause the replica sequence to evolve toward the sought-after minimum energy path (MEP). One of these, the longitudinal component of spring force that connects adjacent replica states is given by:

\begin{equation}
    F_i^S = [k(\mathbf{R_{i+1}}-\mathbf{R_i})-k(\mathbf{R_i}-\mathbf{R_{i-1}})]\cdot \hat{\tau}\hat{\tau}
\end{equation}

where $\mathbf{R_i}$ represents the position of the $i^{th}$ replica in the energy landscape, \textit{k} is the spring constant, and $\hat{\tau}$ is the "longitudinal" unit vector, parallel with the "spring" at replica $i$.

The other, lateral force component is perpendicular to the "spring", exerted by the gradient of the energy surface. As such, this force component nudges the spring towards the MEP. In our SolDeg platform, the energy surface was computed with the Si GAP.

\begin{equation}
    F_i^V = -\nabla V(\mathbf{R_i}) + \nabla V(\mathbf{R_i})\cdot \hat{\tau} \hat{\tau}
\end{equation}

where \textit{V} is the potential energy landscape. The advantage of the NEB method over a standard elastic band method is that artifacts involved with the band cutting corners off the MEP are not a problem for the NEB method.

Our NEB simulations were performed in LAMMPS using the Si GAP \cite{PRX}. We used 32 replicas for each simulation. In our simulations we kept the non-heated atoms fixed: only the heated atoms were allowed to move replica-by-replica. The energy stopping tolerance was $10^{-6}$ (unitless), and the force stopping tolerance was $10^{-6}$ eV/\AA. The simulation timestep was 10 fs. We used the \textit{fire} minimization algorithm, a damped dynamics method with a variable time step \cite{Bitzek}. 

\begin{figure}[h!]
\begin{center}
\includegraphics[width=\columnwidth]{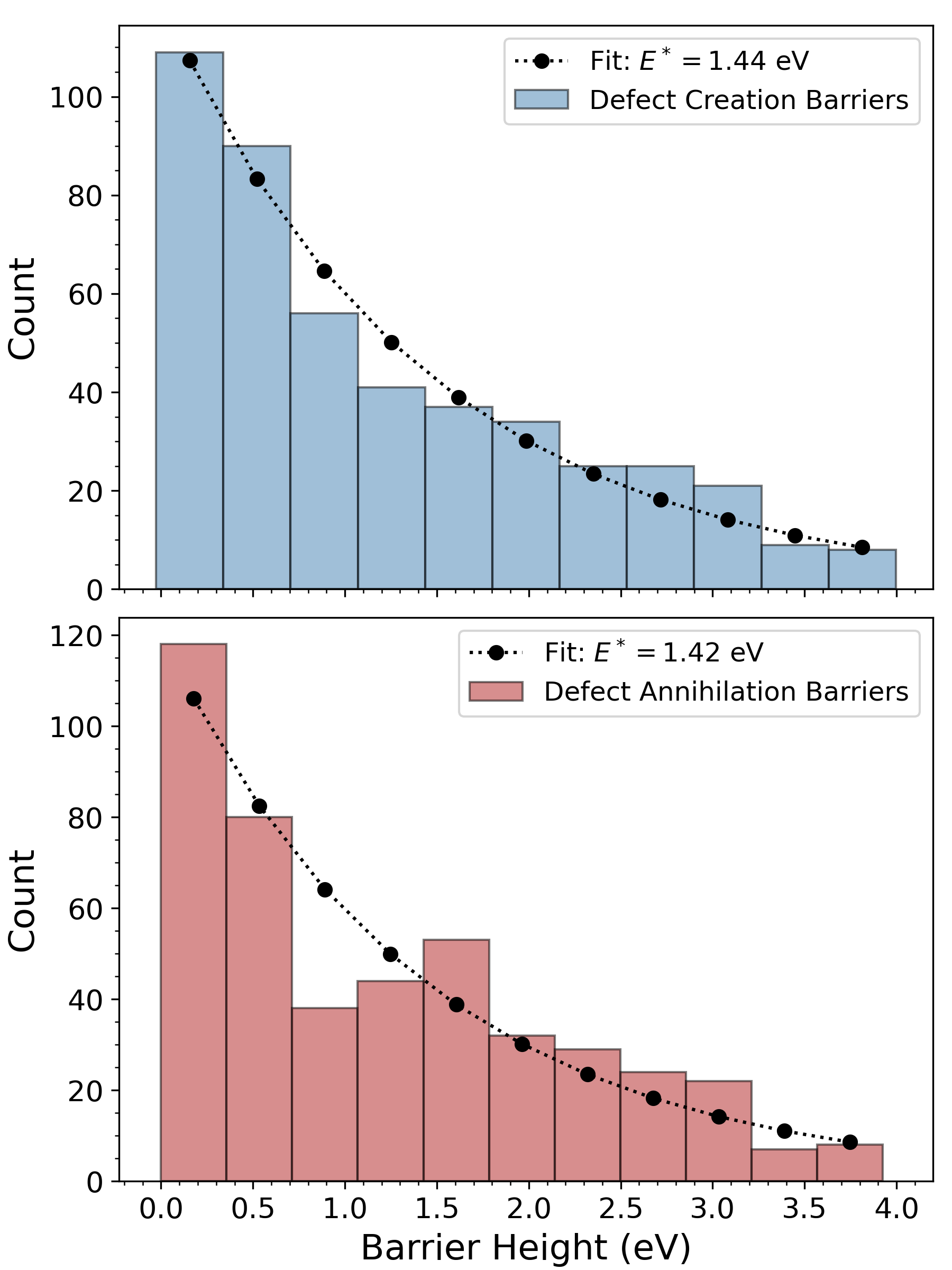}
\end{center}
\caption{Energy barrier distributions for: (a) defect creation; and (b) defect annihilation. Black symbols and lines: exponential fit with: $P(E)=(1/E^*) \textrm{exp}(-E/E^*)$}
\label{barrier_dist}
\end{figure}

The results of our NEB calculation are shown in Fig.~\ref{barrier_dist}. The distribution of the barriers for the defect creation processes is shown in panel (a), while the distribution of the barriers for defect annihilation processes (a reverse transition across the defect creation barrier) is shown in panel (b). Because of the similarity of the two distributions, both the creation and the annihilation processes will impact the time evolution of the defect density. Importantly, Fig.~\ref{barrier_dist} reveals an extremely broad distribution of barriers, from meV to 4 eV. Such an extremely broad barrier distribution is the hallmark of glassy phenomena, and is the driving force behind the defect density growing not only on microscopic time scales but also on the time scale of years.

\subsection{Determining the Temporal Evolution of the Defect Density N(t) From the Barrier Distribution}

In the last stage of the SolDeg platform, we determined the temporal evolution of the defect density from microscopic times scales to 20 years, the standard length of solar cell performance guarantees.

In order to determine the defect density as a function of time, we turned to kinetic Monte Carlo methods. We begun by creating samples with 20,000 individual two-state "clusters" that each could transition from a non-defected state to a defected state by overcoming a defect creation energy barrier, and transition from a defected state to a non-defected state by overcoming a defect annihilation energy barrier. For each cluster transition, the energies for these creation and annihilation processes were drawn from the two barrier distributions determined in the previous section. We eliminated artificial fluctuations induced by the discrete binning of the barriers by representing the distributions with their smooth fitted forms, as shown in Fig. \ref{barrier_dist}. The clusters transition over the barriers by thermal activation, with an associated rate of

\begin{equation}
    \Gamma = \Gamma_0 e^{\frac{-E}{kT}},
\end{equation}

where $\Gamma_0$ is a characteristic attempt frequency of the cluster to overcome its energy barrier, taken here to be $10^{10}$ $\textrm{s}^{-1}$. These rates are calculated for each cluster and summed to determine the "total rate", $\Gamma_{tot}$. Next, an event is randomly selected from the possible pool of events, with the probability of selecting event $i$ being equal to $P(i) = \frac{\Gamma_i}{\Gamma_{tot}}$. The time is then moved forward according to $\Delta t = \frac{-ln(r)}{\Gamma_{tot}}$, where r is a random number. This is equivalent to sampling a Poisson waiting time distribution.

The above described method becomes computationally prohibitive when the phenomena of interest are rare events, with rates of occurrence that are several orders of magnitude smaller than typical events. Not only are these rare events exceptionally unlikely to be chosen by the KMC algorithm, but the number of simulation steps needed to evolve the simulation time far enough to see the rare events will be impossibly large. Using our base kinetic Monte Carlo algorithm, without any acceleration efforts, a million simulation steps only evolve the simulation time by one-hundredth of a second. This is completely inadequate to determine degradation that occurs on the scales of months or years. 

Our solution to this problem was to implement the "accelerated super-basin kinetic Monte Carlo" (AS-KMC) algorithm \cite{AS_KMC}. The AS-KMC method adds the extra algorithmic step of checking whether any of the events that are part of a "super-basin" have been executed a pre-specified number of times. For such events, the AS-KMC increases the barrier height, thereby lowering their rate of occurrence. In conventional terms, a super-basin consists of clusters which are linked to each other by high-frequency events but are separated from the surrounding energy landscape by one or more high barriers, making the frequency of the escape from the basin dramatically lower. AS-KMC avoids getting stuck in a single super basin by boosting the probability of the system overcoming these high barriers. In our implementation, as the clusters are independent from each other, a superbasin is the set of fast transitions over the low energy sector of the barrier distribution for each cluster. The AS-KMC method increasing the barrier height in the low energy sector can be thought of as integrating out the fast degrees of freedoms in a renormalization group sense, thereby mapping the problem to a scaled problem where the slower transitions over the higher energy barriers are the typical processes. The formalism of our implementation of the AS-KMC method is as follows. Every N times that an event occurs, the transition rate $\Gamma$ of that event is reduced by a factor $\alpha$, such that $\Gamma' = \Gamma/\alpha$. Here $\alpha$ is taken as:

\begin{equation}
    \alpha = 1+ (\textrm{N}\,\delta) /|\ln(\delta)|.\,\,
\end{equation}

where $\delta$ is the magnitude of the relative error in the new probability of escaping the superbasin once the internal activation barrier has been raised. We chose $N=10$ and $\delta=0.25$. As is clear from the above description of the KMC method of forwarding the time at each executed transition by the inverse of the executed rate, the scaling of the $\Gamma$ rates scales the time itself. The integrating out of the fast degrees of freedom and the rescaling of time together map the model to a "slower transitions only" model. As this integrating out and rescaling is repeated many times over, transitions over all time scales are properly accounted for by this AS-KMC method.

\begin{figure}[h!]
\begin{center}
\includegraphics[width=\columnwidth]{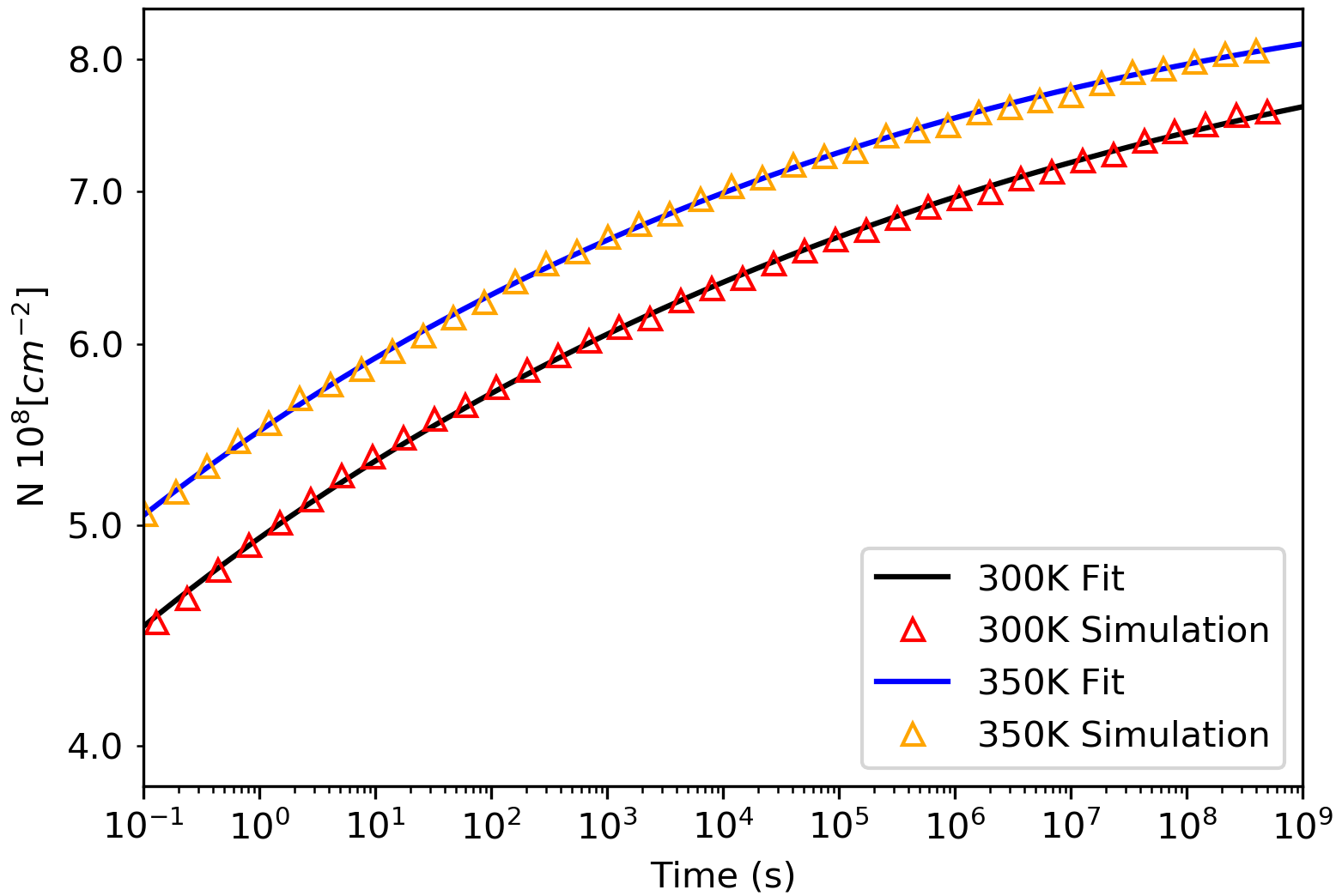}
\end{center}
\caption{Defect density N(t) as a function of time. Defect saturation density was chosen as $N_{sat}=1x10^{9}$ $\textrm{cm}^{-2}$. Orange: defect generation by AS-KMC at $T$=300K. Red: accelerated defect generation by AS-KMC at $T$=350K. Blue and green: Stretched exponential fits to AS-KMC results with stretching exponent $\beta (300 \textrm{K}) = 0.019$, and $\beta (350 \textrm{K}) = 0.022$.}
\label{N(t)}
\end{figure}

We reduced the noise by simulating the AS-KMC dynamics for 64 samples of 20,000 clusters each, and finally by averaging the results. Fig.~\ref{N(t)} shows the time dependent defect density $N(t)$, determined by this method. The minimal fluctuations of $N(t)$ are representative of the effective error bars and thus show that the above approach averaged out the fluctuations very efficiently. The AS-KMC dynamics was performed for samples with temperature at $T$=300K (red), and $T$=350K (orange), in order to simulate defect generation at ambient temperatures and with standard accelerated testing protocols at elevated temperatures, as described below in detail.

In order to develop an analytic model and understanding for these simulation results, we recall that systems that exhibit very slow dynamics are often thought of as glassy systems with a broad distribution of energy barriers $P(E)$\cite{Amir-2011}. In general, the distribution of energy barriers $P(E)$ can be translated into a distribution of "barrier crossing times" $P(\tau)$, and then coupled rate equations can be written down for defect creation and defect annihilation. The expectation value of cluster transition rates at time $t$ can be determined by integrating over the barrier crossing time distribution $P(\tau)$ up to $t$ which turns out to be time dependent instead of the usual constant rates, typical for well defined transition energies.\cite{Freitas2014,Johnston2008} This rate equation for the defect density with time dependent rates can then be solved for $N(t)$. 

The specific time dependence of $N(t)$ depends on the functional form of the energy barrier distribution $P(E)$. As seen in Figs. \ref{barrier_dist}a-b, our $P(E)$ distributions can be well-fitted with an exponential, $P(E) = (1/E^*) \textrm{exp} (-E/E^*)$, with $E^*=1.42$ eV for barrier creation and $E^*=1.44$ eV for barrier annihilation. Following the above steps for an exponential energy barrier distribution yields a stretched exponential time dependence \cite{StretchedExponential1, StretchedExponential2, StretchedExponential3}:

\begin{equation}
    N(t)=N_{sat}\left( 1-\exp{\left[-(t/\tau_0)^{\beta}\right]}\right),
\end{equation}

where $\beta=k_{B}T/E^*$, and $\tau_0$ is a short time cutoff. It is important to emphasize that here $\beta$ was not a fitting parameter. Once we determined $E^*$ from the $P(E)$ we computed earlier (Figs. \ref{barrier_dist}a-b), this fixed the value of $\beta$. $\beta$ being fixed makes it all the more remarkable that we were able to fit $N(t,300\textrm{K})$ {\it over ten orders of magnitude} with $\beta (300\textrm{K})=0.019=k_{B}*300\textrm{K}/E^*$ since $E^*=1.43$ eV, the average of the defect creation and defect annihilation energy scales; and analogously, fit $N(t,350\textrm{K})$ with $\beta (350 \textrm{K})=0.022$. 

For completeness we note that we obtained very good fits setting $\tau_0$ with $1/\Gamma_0$, but our fits improved by using shorter $\tau_0$ cutoff values. Developing a physical interpretation for the best $\tau_0$ is left for a later paper. Further, forcing power law or near-flat fits on $P(E)$ predicted power law and logarithmic time dependencies. Such forms can achieve reasonable fits for $N(t)$ over 2-4 orders of magnitude in time, but as the fitting range was extended, the stretched exponential fit produced the singularly best fit, and thus we conclude that the exponential for for $P(E)$ is the natural choice.

A stretched exponential time dependence was reported for the recombination lifetime at a-Si:H/c-Si interfaces\cite{DeWolf_2008} before, as well as related aging phenomena \cite{Amir-2011}. The measurement "annealing" temperature was T=450K, and the $\beta$ exponent assumed values in the 0.29-0.71 range. Accordingly, the characteristic energy scale $E^*$ of the barrier distribution that controlled this time dependence was in the range of $E^*=50-125$ meV, an order of magnitude smaller than the Si defect energies that control the time evolution in this paper. We agree with the conclusion of the authors of Ref. \citenum{DeWolf_2008}: their time dependence was probably controlled by hydrogen diffusion. Hydrogen was not considered in our model. 

The main messages of Figs. \ref{barrier_dist}a-b and Fig.~\ref{N(t)} are as follows.

(1) It has been customary to think about degradation processes in Si solar cells as being controlled by chemical bonds with well-defined energies, at most with a narrow distribution. But our simulations of realistic a-Si/c-Si stack interfaces show that the bond energies of a large fraction of the Si atoms, especially those close to the interface, are weakened by stretching and twisting, many to the point of being broken. Therefore, the defect generation is controlled by a broad distribution of energy barriers instead of a narrow one. One is led to the conclusion that the solar cell degradation needs to be described in terms of such wide energy barrier distributions.

(2) We developed the SolDeg platform to answer the above need. SolDeg is capable of connecting the fast atomic motions that control defect structures and play out on the femtosecond time scale, with the slow, glassy transitions controlled by the wide distribution of energy barriers that take place over time scales up to gigaseconds, the order of 20 years. The ability of the SolDeg platform to bridge these 24 orders of magnitude in time makes it a uniquely powerful tool for a comprehensive study of defect generation in a-Si/c-Si stacks.

(3) We have shown that a simple, stretched exponential analytical form can successfully describe defect generation over an unparalleled range of ten orders of magnitude in time. This analytical form may turn out to be quite useful for the analysis of experimental degradation studies.

(4) As far as numerical values are concerned, the defect generation rate in the first month (starting from $10^5$ seconds, about one day) is $\approx 1.5\%$/month that slows to 4\%/year for the first year. This defect generation rate will be used to connect our theoretical work to experimental data as follows. The published NREL measurements capture degradation of fielded HJ a-Si/c-Si solar cells in terms of $V_\mathrm{oc}$, the open circuit voltage, with the result of 0.5\%/year in relative terms\cite{DelineChris2018SHSF}. The well-known relation connecting $V_\mathrm{oc}$ to the defect density reads:

\begin{equation}
V_\mathrm{oc} =  \frac{kT}{q} \ln{\left(\frac{J_L}{J_0}\right)} ,
\end{equation}

here $J_0$ is the dark saturation current, proportional to the defect density $N(t)$, and $J_L$ is the light current. One expects that the primary driver of the performance degradation is the defect density, $N(t)$, thus, we can capture the degradation in natural, relative terms as

\begin{equation}
\frac{1}{V_\mathrm{oc}}\frac{dV_\mathrm{oc}}{dt} = \frac{-1}{\ln(J_L/J_0)} \frac{1}{N} \frac{dN}{dt}.
\end{equation}

Using relevant values for $J_L$ and $J_0$ reveals that $ln(\frac{J_L}{J_0})$ is realistically around 20, thus the 4\%/year relative defect density growth rate gives a relative degradation rate for $V_\mathrm{oc}$ of about 0.2\%/year. On one hand, it is a reassuring validation of the quantitative reliability of the SolDeg platform that our $V_\mathrm{oc}$ degradation rate came out to be comparable to the 0.5\%/year change observed in fielded HJ modules\cite{DelineChris2018SHSF}. On the other hand, the fact that our calculated degradation rate is notably lower than the observed value is consistent with the physical expectation that in commercial HJ cells the silicon is hydrogenated, and hydrogen migration is expected to be a primary driver of the defect generation. Further, exposure to illumination also enhances defect generation. The fact that our work has not included hydrogen or illumination yet comfortably accounts for the computed degradation rate being lower than the observed one. Finally, it is noted that while some published experiments report the above steady degradation rate of 0.5\%/year over 7 years \cite{DelineChris2018SHSF}, other, shorter time studies report a strongly slowing degradation \cite{DeWolf_2008}. Our results are consistent with the latter, and thus we think that comparison with experiments should be done in terms of a full time dependence of $N(t)$ or $V_\mathrm{oc}$. At any rate, the natural next step for the development of the SolDeg platform is to include hydrogen and illumination. This demanding work is already ongoing and will be reported soon. 

\begin{figure}[h!]
\begin{center}
\includegraphics[width=\columnwidth]{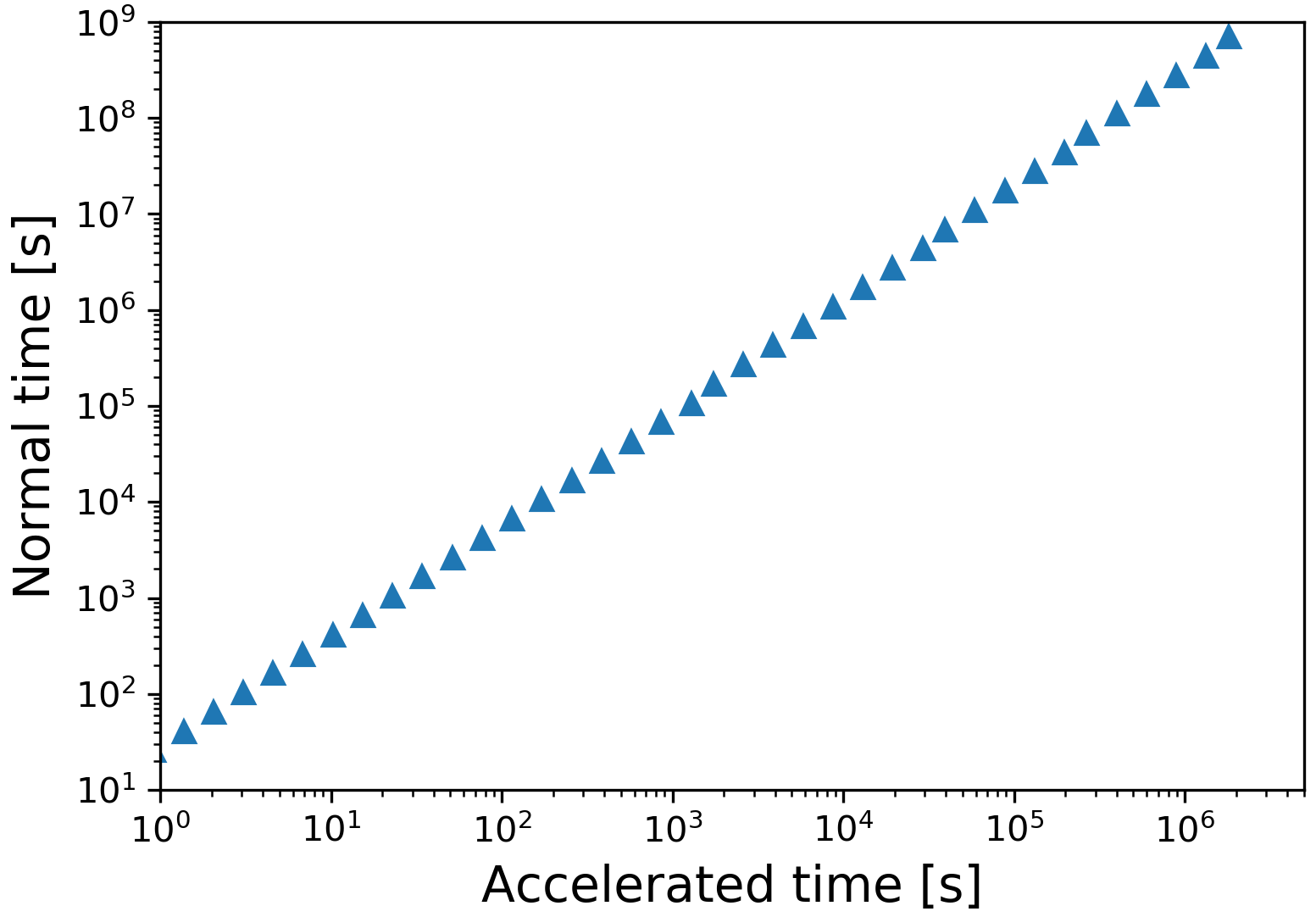}
\end{center}
\caption{Time Correspondence Curve, translating accelerated degradation time to degradation time at standard temperature.}
\label{TCC}
\end{figure}

(5) The power of SolDeg can be further demonstrated by developing a quantitative guide to calibrate the widely used accelerated testing protocols of solar cells. Fig.~\ref{N(t)} also shows the accelerated growth of the defect density in a HJ stack at the elevated temperature of $T$=350K. The two simulations were started with the same defect density at $t$=0: $N$($T$=300K, $t$=0)=$N$($T$=350K, $t$=0). Visibly, the $T$=300K and $T$=350K curves largely track each other: the difference is that $N$($T$=350K, $t$) reaches the same defect densities as $N$($T$=300K, $t$) at shorter times. This is why week-long accelerated testing can capture year-long defect generation under ambient/fielded conditions.

To turn this general observation into a quantitatively useful calibration tool, we created the Time Correspondence Curve (TCC). The TCC connects the times of accelerated testing with those times of normal, ambient degradation that produce the same defect density. In formula: TCC plots the $t_\mathrm{accelerated}$ -- $t_\mathrm{normal}$ pairs for which $N$($T$=350K, $t_\mathrm{accelerated}$) = $N$($T$=300K, $t_\mathrm{normal}$). Fig.~\ref{TCC} shows the resulting TCC. For example, the TCC shows that $t_\mathrm{accelerated}$=$10^6$ seconds of accelerated testing approximately generates the same density of defects as $t_\mathrm{normal}$=$10^8$ seconds or normal degradation. In general, the TCC was created by taking horizontal slices across the two curves of Fig.~\ref{N(t)} to find the corresponding pairs of times that generated the same defect density. 

Even a cursory observation reveals that the TCC grows linearly on the log-log plot, i.e. as a power law over an extended, experimentally relevant time period:

\begin{equation}
t_\mathrm{accelerated} \propto t_\mathrm{normal}^{\nu}
\end{equation}

where $\nu=0.85\pm0.05$, an unexpected scaling relation with an unexpected precision. Just like in Fig. ~\ref{N(t)}, this scaling relation is observed over the most remarkable ten orders of magnitude in time. Establishing such simple and practical correspondence relations can be a very helpful product of the SolDeg platform that can be widely used for calibrating accelerated testing protocols. Remarkably, the above-developed description in terms of an exponential $P(E)$ that led to an stretched exponential $N(t)$ that explained the results over ten orders of magnitude, also gives a straightforward explanation for this scaling relation. Direct observation of the stretched exponential formula reveals that $N$($T$=350K, $t_\mathrm{normal}^{300/350}$) = $N$($T$=300K, $t_\mathrm{normal}$), i.e. the stretched exponential form not only explains the existence of the scaling form of TCC, but makes a prediction for $\nu$:

\begin{equation}
\nu= T_\mathrm{normal}/T_\mathrm{accelerated}=300/350=0.85,
\end{equation}

which is exactly the exponent what the direct analysis of the TCC determined. These considerations provide a remarkably self-consistent and powerful tool set to analyze degradation processes. 

\begin{figure*}[h!]
\begin{center}
\includegraphics[width=\textwidth]{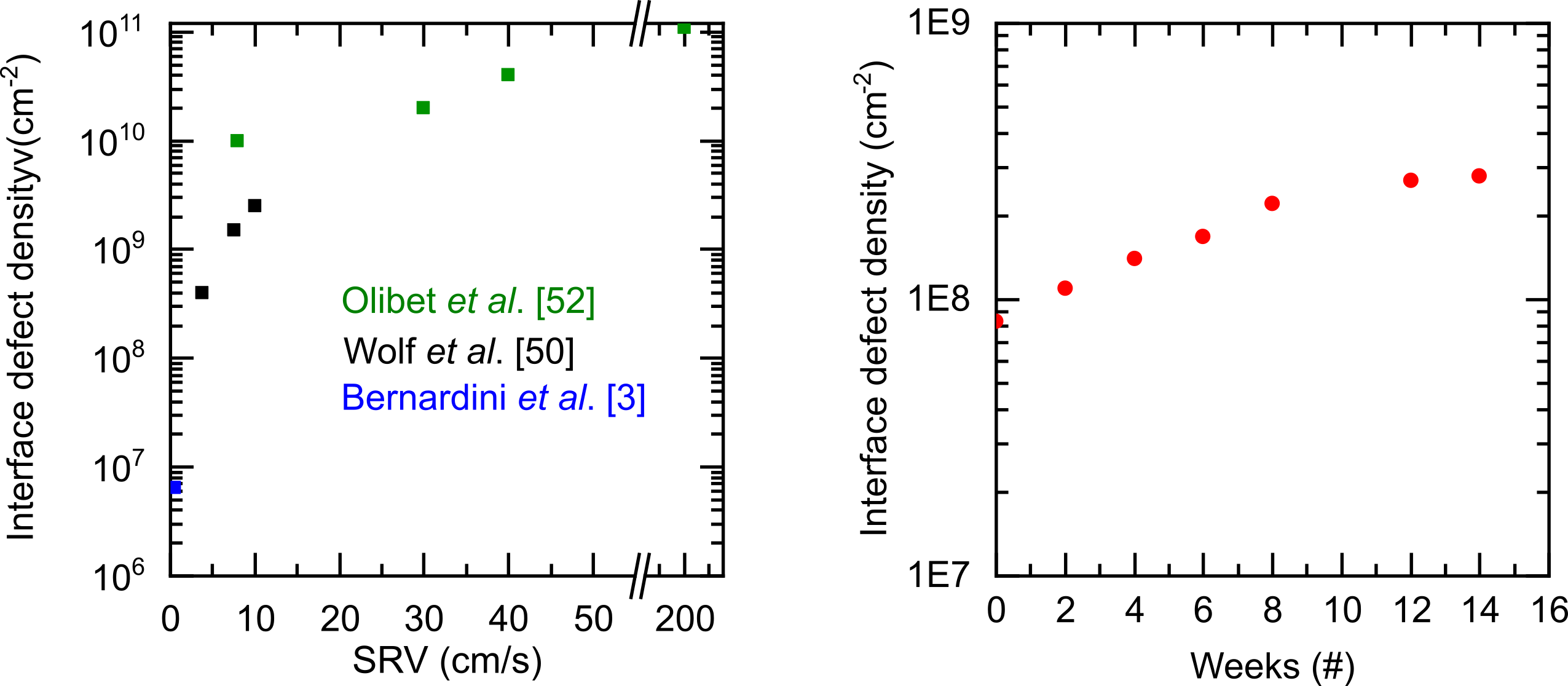}
\end{center}
\caption{(left) Interface defect density and effective surface recombination velocity data reported in the literature for a-Si:H/c-Si interface at the time of deposition or after annealing. (right) Interface defect density measured in our a-Si:H/c-Si stacks, stored in the dark under standard ambient condition.}
\label{expFig}
\end{figure*}

\section{Experimental Studies of Degradation of a-Si:H/c-Si stacks}

In this section, we explore the correspondence between our simulations and experiments on a-Si:H/c-Si heterojunction structures. Fig.~\ref{expFig}(a) shows that for the interface defect density, a wide range of values have been reported in the literature \cite{Bonilla2017,Olibet2007,Olibet2010,Schulze2010}. This unusually wide range is caused by many different factors, such as the different methods and protocols employed for depositing a-Si, the level of cleanliness of c-Si wafer before deposition, substrate morphology, orientation of c-Si, microstructure of the a-Si film, hydrogen content, bonding in a-Si, and storing conditions of samples \cite{Stuckelberger2013, Holovsky2020, Schulze2011, Smets2003}. However, there are not many studies available correlating the impact of such differences to long-term stability of a-Si/c-Si interface. For completeness, in Fig.~\ref{expFig}(a) we also summarize the corresponding surface recombination velocities (SRV) from the cited papers. Disappointingly, very few of these papers have analyzed the time dependence of the interface defect density and that of the SRV. Therefore, it is difficult to draw lessons from these papers for the long term performance degradation of heterojunction structures.

Driven by these considerations, we have set out to fabricate our own a-Si:H/c-Si heterojunction structures to measure the long time evolution of the defect density and SRV. In order to experimentally isolate the processes associated with the time evolution of interface defect density and SRV from the processes occurring in other layers, and to be able to access these quantities with direct measurements, we created test structures that were simpler than Si HJ solar cells. This is to isolate the changes happening at the interface and in the film over time from the influence of other layers present in the cell. 

These simpler structures, or stacks, were comprised of c-Si of varying thickness (160–260 $\mu$m), passivated on both sides with hydrogenated intrinsic a-Si (a-Si:H(i)). We used double-side polished float zone (FZ) quality \textit{n}-type c-Si wafers with (100) crystal orientation, 2.5 $\Omega$cm resistivity and initial thickness of $\sim$275 $\mu$m. These wafers went through rigorous surface cleaning before deposition of 50 nm of a-Si:H(i) on both sides. Complete details about cleaning protocol and deposition conditions are found in \cite{bernardini2018}. After deposition, these samples were annealed at 280 $\degree$C for 30 mins in air in a muffle furnace. Then we measured injection-dependent effective minority carrier lifetime ($\tau_{eff}$) on these samples at temperatures between 30 and 230 $\degree$C using the WCT-120TS tool from Sinton Instruments. Data was collected in transient mode due to the long lifetime of the samples. We performed linear fits for 1/$\tau_{eff}$ vs 1/\textit{W} data at each injection level to obtain temperature- and injection-dependent SRVs from their slopes. Here \textit{W} is thickness of the c-Si. As a final step we extracted the interface defect density by fitting the amphoteric defect model proposed by Olibet \textit{et al.} to the SRV \textit{vs} temperature data at different injection levels. The input parameters, along with their best-fit values to the model, were the interface charge density (\textit{Q} = $-1.3 \times 10^{11} \,\mathrm{cm}^{-2}$), the neutral electron-to-hole capture cross-section ratio ($\frac{\sigma^o_n}{\sigma^0_p} = \frac{1}{20}$), and the charged-to-neutral capture cross-section ratio ($\frac{\sigma^+_n}{\sigma^o_n} = \frac{\sigma^-_p}{\sigma^o_p} = 500$) with $\sigma^o_p = 10^{-16} \,\mathrm{cm}^{-2}$.

The resulting interface defect density values over time for the samples stored in dark and ambient conditions are shown in Fig.~\ref{expFig}(b). We found that the magnitude of the fitted interface charge density remained the same over time, indicating no change in field effect passivation \cite{DeWolf_2008}. However, the interface defect density increased with time at the rate of \textit{dN/dt}$=5.6\times10^7 \mathrm{cm}^{-2}$/month. This translates to a rate of increase of $\textit{(1/N) dN/dt}=68\%$/month in relative terms for the first 3.5 months. Similar results have been reported previously by Bernardini et al., where the defect density increased from $(6.5\pm0.5)\times10^5 \mathrm{cm}^{-2}$ to $(5.5\pm1.5)\times10^7 \mathrm{cm}^{-2}$ over a period of 28 months for samples stored in dark, ambient conditions \cite{BertoniMarianaI.2018IitD}. This translates to a rate of $17\%$/month in relative terms. While our study found a higher rate in the initial 3.5 months, we expect the defect generation rate to slow down considerably as time progresses, and converge to the results of Ref.\cite{BertoniMarianaI.2018IitD}. The fact that the defect generation slows down was clearly established by our simulations as well, as shown in Fig.~\ref{N(t)}, where we found that the rate of change in the early months is about 1.5\%/month, slowing to an overall rate of 4\%/year for the first year. 

To make connection to previous results, we fitted our experimental data with a stretched exponential. Holovsky. et al. reported such fits with stretching exponent $\beta$ in the 0.3-0.7 range, depending on deposition temperatures\cite{DeWolf_2008}. Our data are consistent with the stretching exponent in this range. However, the relatively limited temporal range of the data does not conclusively exclude $\beta$ values outside this range either. We continue taking data that will narrow the range of the stretching exponent.

We note that the initial defect density in our samples was lower than those reported for as-deposited films in Fig.~\ref{expFig}(a) by at least an order of magnitude. This could be due to the difference of the deposited a-Si:H film in terms of the crystallinity, hydrogen content, hydrogen bonding configuration and void fraction\cite{Wu}. Whatever the reason may be, the notably low defect density is a compelling indicator for the high quality of our a-Si deposition protocol.

It is noted, of course, that the degradation rate observed in our test structures shown in Fig.~\ref{expFig}(b) is not expected to directly correspond to the degradation rate observed in complete Si HJ cells. This is due to the additional layers present in Si HJ cells on top of a-Si:H(i) which may efficiently suppress the migration of hydrogen away from the a-Si:H/c-Si interface, as well as prevent oxidation of the a-Si:H(i) layer. 

\section{Conclusions}

In this paper we reported the development of the SolDeg platform for the study of heterojunction solar cell degradation. SolDeg layers several techniques on top of each other, in order to determine the dynamics of electronic defect generation on very long time scales. The first layer of SolDeg was to adapt LAMMPS Molecular Dynamics simulations to create a-Si/c-Si stacks. Our simulations used femtosecond time-steps. For the interatomic potential, we used the machine-learning-based Gaussian approximation potential (GAP). Next, we optimized these stacks with density functional theory calculations. In SolDeg's next layer we created about 1,500 shocked clusters in the stacks by cluster blasting. We then analyzed the just-generated shocked clusters by the inverse participation ratio (IPR) method to conclude that cluster blasting generated electronic defects in about 500 of the 1,500 shocked clusters. Next, we adapted the nudged elastic band (NEB) method to determine the energy barriers that control the creation and annihilation of these electronic defects. We performed the NEB method for about 500 shocked clusters on our way to determine the distribution of these energy barriers. A simple exponential form gave a good fit for $P(E)$.  Finally, we developed an accelerated super-basin kinetic Monte Carlo (AS-KMC) approach to determine the time dependence of the electronic defect generation, as controlled by the broad energy barrier distribution.

Our main conclusions were as follows. (1) The degradation of a-Si/c-Si heterojunction solar cells via defect generation is controlled by a very broad distribution of energy barriers, extending from the scale of meV to 4 eV. (2) We developed the SolDeg platform that can track the microscopic dynamics of defect generation $N(t)$ from femtoseconds to gigaseconds, over 24 orders of magnitude in time. This makes SolDeg a uniquely powerful tool for a comprehensive study of defect generation in a-Si/c-Si stacks and solar cells. (3) We have shown that a simple, stretched exponential analytical form can successfully describe the defect generation $N(t)$ over ten orders of magnitude in time. (4) We found that in relative terms $V_\mathrm{oc}$ degrades at a rate of 0.2\%/year over the first year. It is a reassuring validation of the quantitative reliability of the SolDeg platform that our $V_{oc}$ degradation rate came out to be comparable to the 0.5\%/year change observed in fielded HJ modules \cite{DelineChris2018SHSF}. The difference is most likely attributable to the SolDeg platform not yet including hydrogen and illumination. The project to include both has already started and will be reported in a later publication. (5) Further, we developed the Time Correspondence Curve to calibrate and validate accelerated testing of solar cells. This TCC connects the times of accelerated testing with those times of normal, fielded degradation that produce the same defect density. Intriguingly, we found a compellingly simple scaling relationship between accelerated and normal times $t(\mathrm{accelerated}) \propto t(\mathrm{normal})^{0.85}$, which can be used to calibrate accelerated testing protocols, making it a more quantitative assessment tool. (6) We ourselves also carried out experimental work on defect generation in a-Si/c-Si HJ stacks. We found that the degradation rate was high on the short, initial time scales, but slowed considerably at longer time scales. A possible explanation is that our samples had unusually low initial defect densities, in which case hydrogen diffusion may generate defects more efficiently.

We plan to continue this project by incorporating hydrogen and illumination into the SolDeg platform and determine the dynamics of the defect generation anew. We will validate and calibrate the improved SolDeg platform by a rigorous comparison to our experimental data. Once the driving forces of defect generation and degradation are reliably captured and characterized by the SolDeg platform and by our experiments, we plan to develop strategies to mitigate these degradation processes.

\section*{Conflicts of interest}
There are no conflicts to declare.

\section*{Acknowledgements}
We acknowledge useful and inspiring discussions with Ariel Amir, Steven Hegedus, Ron Sinton, Arno Smets, David Strubbe, and Christopher Thomson. This research was supported by DOE SETO grant DE-EE0008979.

\section*{Software Availability}

The GAP suite of programs is freely available for non-commercial use from www.libatoms.org. The Quantum Espresso software package is freely available from www.quantum-espresso.org. The LAMMPS software package is freely available from lammps.sandia.gov. 

\balance

% \bibliography{bibliography.bib}
% \bibliographystyle{rsc}
\providecommand*{\mcitethebibliography}{\thebibliography}
\csname @ifundefined\endcsname{endmcitethebibliography}
{\let\endmcitethebibliography\endthebibliography}{}

\end{document}